\documentclass{aa}  

\usepackage{graphicx}
\usepackage{txfonts}
\usepackage[colorlinks=true,linkcolor=blue,citecolor=blue,urlcolor=blue]{hyperref}

\usepackage{amsmath,amssymb,bm}
\usepackage{BOONDOX-cal}
\usepackage{subcaption}
\usepackage{booktabs,multirow,makecell,array,threeparttable,tabularx}
\usepackage{siunitx}
\usepackage{xcolor}
\usepackage{float}
\usepackage{adjustbox}

\newcolumntype{Y}{>{\centering\arraybackslash}X}
\newcolumntype{L}[1]{S[table-format=#1]}

\begin{document}

\title{Prospects for cosmological research using hundred-meter-class radio telescopes:
21-cm intensity mapping survey strategies with QTT, JRT, and HRT}

\titlerunning{21-cm IM with QTT, JRT, and HRT}
\authorrunning{J.-D. Pan et al.}

\author{
Jun-Da Pan\inst{1}
\and
Yichao Li\inst{1}
\and
Guo-Hong Du\inst{1}
\and
Tian-Nuo Li\inst{1}
\and
Xin Zhang\inst{1,2,3,}\thanks{Corresponding author: Xin Zhang, e-mail:\texttt{zhangxin@neu.edu.cn}}
}

\institute{
Key Laboratory of Cosmology and Astrophysics (Liaoning), College of Sciences,
Northeastern University, Shenyang 110819, China
\and
Key Laboratory of Data Analytics and Optimization for Smart Industry (Ministry of Education),
Northeastern University, Shenyang 110819, China
\and
National Frontiers Science Center for Industrial Intelligence and Systems Optimization,
Northeastern University, Shenyang 110819, China
}

\abstract{
Understanding dark energy requires precision measurements of the expansion history of the universe and the growth of large-scale structure. The 21\,cm intensity mapping (21\,cm IM) technique enables rapid large-area surveys that can deliver these measurements. China is constructing three hundred-meter-class single-dish radio telescopes, including the QiTai 110\,m Radio Telescope (QTT), the 120\,m Jingdong Radio Telescope (JRT), and the 120\,m Huadian Radio Telescope (HRT), whose designs are well suited for 21\,cm IM cosmology. We use a Fisher-to-MCMC forecasting framework to evaluate the baryon acoustic oscillations / redshift space distortions (BAO/RSD)  measurement capabilities of QTT, JRT, and HRT and propagate them to dark-energy constraints in the $w_0w_a$CDM model. Our results show that achieving a redshift coverage up to $z_{\mathrm{max}} = 1$ is crucial for fully realising the potential of hundred-meter-class single-dish telescopes for 21\,cm cosmology. If all three telescopes carry out 21\,cm IM surveys over the same redshift range up to $z_{\mathrm{max}}=1$ and combine their BAO/RSD measurements, QTT+JRT+HRT yield $\sigma(w_0)=0.094$ and $\sigma(w_a)=0.487$, providing tighter constraints than DESI DR2 results.
}


\maketitle


\section{Introduction}

The discovery of the accelerated expansion of the universe challenges our understanding of cosmic composition and gravity \citep{SN1999,SN1998}. The cause may involve an unknown energy component with unusual physical properties, known as dark energy, or it may indicate a modification of general relativity on cosmological scales. To uncover the nature of this phenomenon, high-precision measurements of the expansion history of the universe and the evolution of large-scale structure (LSS) are essential \citep{Frieman:2008sn,Weinberg:2013agg,Joyce:2016vqv}. Baryon acoustic oscillations (BAO) provide a standard ruler imprinted by sound waves in the early universe and are widely used to trace the expansion history \citep{BAO:1998,BAO:2003pu,BAO:2003rh}. At present, BAO measurements are obtained by mapping galaxy distributions in large optical surveys, particularly with the second data release (DR2) from the Dark Energy Spectroscopic Instrument (DESI) marking the onset of a new era in high-precision BAO measurements~\citep{DESI1}. The DESI DR2 BAO data, combined with type Ia supernova and cosmic microwave background data, provide more stringent constraints on the dark energy equation of state (EoS) parameters, revealing a $\sim 4\sigma$ deviation from the $\Lambda$CDM paradigm and suggesting a preference for dynamical dark energy within the $w_0w_a$CDM model~\citep[see][]{DESI1}, prompting extensive discussion (e.g., \citealt{Giare:2024gpk, Li:2024qso, Ye:2024ywg}).
\nocite{Du:2024pai,Li:2024qus,Giare:2024smz,Jiang:2024xnu,Li:2025owk,Pan:2025qwy,Yang:2025uyv,Paul:2025wix,Pedrotti:2025ccw,Cai:2025mas,Pang:2025lvh,Wang:2025ljj,Li:2025ops,Li:2025eqh,Jia:2025poj,Adam:2025kve,Li:2025dwz,Luciano:2025dhb,Li:2025htp,Paliathanasis:2025kmg,Du:2025xes,Wu:2025vfs,Hussain:2025uye,vanderWesthuizen:2025rip,Li:2025muv,Yao:2025twv,Zhou:2025nkb,Zhang:2025dwu,Du:2025csv,SanchezLopez:2025uzw,Alam:2025epg,Huang:2025som,Liu:2025myr,Liu:2024yib,Liu:2025mub}

While DESI DR2 has provided the most precise BAO measurements to date,
surveying wider areas and reaching higher redshifts requires longer exposures, more elaborate target selection strategies, and is further constrained by shot noise at low number densities and by fiber assignment constraints \citep{Feldman_1994,Seo_2007,Smith:2018tyi,DESI:2022gle,DESI:2023mkx}. These practical limitations motivate the search for techniques that can rapidly map the LSS of the universe. A more efficient observational approach is 21\,cm intensity mapping (21\,cm IM), which measures the integrated 21\,cm emission from many unresolved galaxies rather than detecting individual sources, with the observed frequency directly determining the redshift.
Therefore, 21\,cm IM enables rapid wide-area surveys and the construction of three-dimensional maps. 
Specifically, the 21\,cm line arises from the ground-state hyperfine transition of neutral hydrogen (\(\mathrm{H\,{\scriptscriptstyle I}}\)), with an emission frequency \(\nu_{\rm emission}=1420.4\,\mathrm{MHz}\) (\(\lambda\approx21\,\mathrm{cm}\)). 
After reionization most neutral hydrogen resides in galaxies, thus 21\,cm IM can trace the galaxy distribution without resolving individual galaxies. 
In practice, 21\,cm IM measures fluctuations in the 21\,cm line brightness temperature, producing \(\mathrm{H\,{\scriptscriptstyle I}}\) maps of large-scale structure that trace the matter distribution on linear scales. 
From these maps one can measure BAO and redshift-space distortions (RSD), thereby constraining the Hubble expansion rate \(H(z)\), the angular diameter distance \(D_\mathrm{A}(z)\), and the growth rate of structure \(f\sigma_8(z)\) (e.g., \citealt{Loeb:2008hg,Mao:2008ug,Lidz:2011dx}
\nocite{Battye:2012tg,Masui_2013,Xu:2014bya,bull2015late,Xu:2017rfo,Yohana:2019ahg,Zhang:2019ipd,Liu_2020,Tramonte:2020csa})

21\,cm IM relies on advanced radio facilities. Current representative instruments include the Five-hundred-meter Aperture Spherical Telescope (FAST)~\citep{nan2011fast, Li:2023zer}, the Green Bank Telescope (GBT)~\citep{Chang:2010jp}, MeerKAT~\citep{2017arXiv170906099S,2021MNRAS.505.3698W,2020MNRAS.tmp.3636L}, the Tianlai experiment~\citep{chen2011radio,chen2012tianlai,2020SCPMA..6329862L,2021MNRAS.506.3455W,2022MNRAS.517.4637P,2022RAA....22f5020S}, and the Canadian Hydrogen Intensity Mapping Experiment (CHIME)~\citep{Newburgh:2014toa}. At present, 21\,cm signals have been detected through cross-correlation with galaxy surveys~\citep{Chang:2010jp, Masui_2013, Anderson:2017ert, cunnington2023h, Tramonte:2020csa, Wolz_2021, 2023HII}, and MeerKAT has reported an auto-correlation detection on Mpc scales~\citep{Paul:2023yrr}. However, the large-scale auto-power spectrum has not yet been directly observed~\citep{switzer2013determination}. Future instruments, including the Square Kilometre Array (SKA)~\citep{dewdney2009ska, santos2015cosmology, bourke2015advancing, 2020, an2022status}, BINGO~\citep{battye2012bingo, wuensche2019bingo}, HIRAX~\citep{Newburgh:2016mwi}, FAST Core Array~\citep{jiangpeng, Pan_2025, Li:2025qkm, Wang:2025ovi}, QiTai 110\,m Radio Telescope (QTT)~\citep{QTT}, 120\,m Jingdong Radio Telescope (JRT)~\citep{JRT:}, and 120\,m Huadian Radio Telescope (HRT) are expected to significantly improve sensitivity and mapping speed.

For different telescope designs, extremely large single-dish telescopes such as FAST provide excellent sensitivity and angular resolution, but their narrow instantaneous field of view limits the mapping speed of wide-area cosmological surveys. In contrast, smaller dishes like the 13.5\,m MeerKAT antennas in single-dish mode offer a wider field of view and faster mapping, yet their low angular resolution smooths out transverse BAO modes at higher redshifts, restricting detections to low-$z$ ranges with limited comoving volume. Hundred-meter-class single-dish telescopes strike a favorable balance between resolution and survey speed. They can resolve transverse BAO modes at relatively high redshifts while covering large sky areas, making them well suited for 21\,cm IM cosmology. Meanwhile, the phased array feed (PAF) technology offers a new way to further enhance mapping efficiency. By deploying a multi-beam receiver array on the focal plane, a PAF can significantly expand the instantaneous field of view without compromising sensitivity, thus greatly improving the survey speed. This technology has already been demonstrated on several telescopes such as ASKAP and GBT, where PAF systems have been installed and tested, showing great potential for wide-area surveys~\citep{PAF1, PAF2}. The combination of hundred-meter-class single-dish telescopes and PAF technology will play an important role in future 21\,cm IM research.
  
Several new hundred-meter-class single-dish radio telescopes are being planned in China. These new instruments will provide important support for 21\,cm IM and LSS studies. First, QTT is a fully steerable 110\,m radio telescope designed to cover a frequency range from approximately 150\,MHz to 115\,GHz and is expected to be equipped with ultra-wideband and PAF \citep{QTT,QTTPAF,QTTP}. This will help expand the redshift range and the scales accessible for BAO measurements. Second, JRT is a fully steerable 120\,m radio telescope located in Jingdong County, Pu’er City, Yunnan Province. It is designed to cover a frequency range of 0.1--10\,GHz and will include wideband receiver \citep{JRT:}. Located at a latitude of approximately \(24^\circ\mathrm{N}\), it offers good sky coverage, making it suitable for long-term, wide-area 21\,cm IM. Third, the HRT is a 120\,m fully steerable radio telescope planned for construction in Huadian City, Jilin Province.

In this work, we forecast BAO/RSD constraints for QTT, JRT, and HRT individually and jointly and propagate them to $\Lambda$CDM and $w_0w_a$CDM parameters; we also examine extending the frequency coverage to $z_{\mathrm{max}}=1$.

For the fiducial cosmology, we adopt the \textit{Planck} best-fit \(\Lambda\)CDM model~\citep{Planck2018}. The parameters are \(H_0 = 67.3\) km\,s$^{-1}$\,Mpc$^{-1}$, \(\Omega_\mathrm{m} = 0.317\), \(\Omega_\mathrm{b} = 0.0495\), \(\Omega_\mathrm{K} = 0\), \(\sigma_8 = 0.812\), and \(n_\mathrm{s} = 0.965\).
\section{21\,cm intensity mapping}
\label{sec2}

In this work, we quantify the observing potential of the telescopes considered here for 21\,cm IM and evaluate the forecasted errors on the BAO/RSD parameters \(\{D_\mathrm{A}(z),\,H(z),\,f\sigma_8(z)\}\). We adopt the semi-analytic Fisher-matrix framework of \cite{bull2015late}, which maps the theoretical 21\,cm signal and instrumental response at the power-spectrum level directly into error forecasts for these quantities.
This approach is computationally efficient, allowing rapid scans over frequency coverage, channel width, system temperature, dish size and beam count, survey area, and total integration time during the design and optimization stage. The resulting BAO/RSD parameter errors are then propagated to cosmological parameters using a Gaussian likelihood explored with Markov Chain Monte Carlo (MCMC), providing an assessment of how instrument design choices impact dark-energy constraints.

\subsection{Signal}
\label{sec2.1}

Following \cite{BAO:2003rh,bull2015late}, we describe the 21\,cm IM signal by the mean $\mathrm{H\,{\scriptscriptstyle I}}$ brightness temperature $\overline{T}_\mathrm{b}$ and the linear Kaiser form. The mean temperature is~\citep[see][]{bull2015late}
\begin{equation}
\overline{T}_\mathrm{b}(z)
= \frac{3}{32\pi}\,
\frac{\mathcal{h} c^3 A_{10}}{k_\mathrm{B} m_\mathrm{p} \nu_{21}^2}\,
\frac{(1+z)^2}{H(z)}\,
\Omega_{\mathrm{H\,{\scriptscriptstyle I}}}(z)\,\rho_{\mathrm{c},0},
\end{equation}
where \(\mathcal{h}\) is Planck's constant, \(A_{10}\) the Einstein coefficient for the 21\,cm transition~\citep{furlanetto2006}, \(k_\mathrm{B}\) Boltzmann's constant, \(m_\mathrm{p}\) the proton mass, and \(\rho_{\mathrm{c},0}\) the present critical density.

The signal covariance in $(\boldsymbol{q},y)$ space is expressed as~\citep{BAO:2003rh,bull2015late}
\begin{equation}\label{eq:cs}
C^\mathrm{S}(\boldsymbol{q},y)
= \frac{\overline{T}_\mathrm{b}^2(z)}{r^2\,r_\nu}\,
\bigl(b_{\mathrm{H\,{\scriptscriptstyle I}}} + f\,\mu^2\bigr)^2\,
\exp\!\left[-k^2 \mu^2 \sigma_{\mathrm{NL}}^2\right]\,
P(k,z),
\end{equation}
with \(\boldsymbol{q}=\boldsymbol{k}_\perp r\), \(y=k_\parallel r_\nu\), \(r\) the comoving distance to redshift \(z\), and \(r_\nu \equiv \left|\mathrm{d}r/\mathrm{d}\nu\right|\) the comoving distance per unit frequency interval.
Here \(b_{\mathrm{H\,{\scriptscriptstyle I}}}\) is the \(\mathrm{H\,{\scriptscriptstyle I}}\) bias, \(f\) the linear growth rate, \(\mu \equiv k_\parallel/k\), \(\sigma_{\mathrm{NL}}=7~\mathrm{Mpc}\) the non-linear dispersion scale, and \(P(k,z)\) the matter power spectrum computed with \texttt{CAMB}~\citep{Lewis_2000}.
This formulation directly links the theoretical signal and the instrumental mapping between \((k_\perp,k_\parallel)\) and \((\boldsymbol{q},y)\), and will be used in the Fisher-matrix forecasts below.

\subsection{Noise}
\label{sec2.2}

The thermal noise covariance for single-dish IM is given by~\citep{bull2015late}

\begin{equation}\label{eq:cn}
C^\mathrm{N}(\boldsymbol{q},y)
=\frac{T_\mathrm{sys}^2}{t_\mathrm{tot}\,\Delta\nu\,N_\mathrm{b}\,N_\mathrm{d}}\,
U_\mathrm{bin}\,B_\perp^{-2}(\boldsymbol{q})\,B_\parallel^{-1}(y),
\end{equation}
where $T_\mathrm{sys}=T_\mathrm{inst}+T_\mathrm{sky}$. 
Here $T_\mathrm{inst}$ is the instrument noise temperature, and $T_\mathrm{sky}$ is the sky brightness temperature dominated by atmospheric and background radio emission, which we approximate as $T_\mathrm{sky}\simeq60~\mathrm{K}\,(\nu/300~\mathrm{MHz})^{-2.55}$. 
$t_\mathrm{tot}$ denotes the total integration time, $N_\mathrm{b}$ the number of beams, $N_\mathrm{d}$ the number of dishes, and $\Delta\nu$ the bandwidth of a redshift bin. The survey-volume factor is $U_\mathrm{bin}=S_\mathrm{area}\,\Delta\tilde{\nu}$, with $\Delta\tilde{\nu}\equiv\Delta\nu/\nu_{21}$ and survey area $S_\mathrm{area}$.
The instrumental responses along the line of sight and transverse directions are
\begin{equation}
B_\parallel(y)=\exp\!\left[-\frac{(y\,\delta\nu/\nu_{21})^2}{16\ln 2}\right],\qquad
B_\perp(\boldsymbol{q})=\exp\!\left[-\frac{(q\,\lambda/D_\mathrm{dish})^2}{16\ln 2}\right],
\end{equation}
where \(\delta\nu\) is the channel width, \(D_\mathrm{dish}\) the dish diameter, and \(\lambda=c/\nu\).

{We note that the thermal-noise model above does not account for a number of instrumental systematics that are known to affect real 21\,cm IM observations, such as $1/f$ noise, beam chromaticity, frequency-dependent sidelobes, polarization leakage, and imperfect polarization calibration. These effects can significantly complicate foreground subtraction and degrade the recovery of cosmological modes, particularly on large scales. Although such systematics are not directly modelled in our forecasts, extensive efforts are underway to characterise and mitigate them using improved instrument design, calibration strategies, and data-driven correction techniques (e.g., \citealp{2020MNRAS.tmp.3636L, Zhang:2021yof, Hu_2021, Gao:2022xdb, Ni:2023ume, Gao:2024nns, Sun:2024ywb, Zhao:2024ita}). The constraints obtained here should therefore be considered optimistic estimates under ideal conditions.}
\subsection{Foregrounds}
\label{sec2.3}

The dominant foregrounds are Galactic synchrotron emission, Galactic free--free emission, and extragalactic point sources. We describe the residual power spectrum after foreground subtraction using the parametric model of \citep{Santos_2005} and \citep{bull2015late},
\begin{equation}
C^{\mathrm{F}}(\boldsymbol{q},y)
= \varepsilon_{\mathrm{FG}}^{2}
\sum_{X} A_{X}
\left[\frac{l_{p}}{2\pi q}\right]^{n_{X}}
\left[\frac{\nu_{p}}{\nu_{i}}\right]^{m_{X}},
\end{equation}
where $\varepsilon_{\mathrm{FG}}$ quantifies the level of residuals: $\varepsilon_{\mathrm{FG}}=0$ corresponds to perfect subtraction and $\varepsilon_{\mathrm{FG}}=1$ to no subtraction. {In this study, we adopt a residual foreground scaling factor of $\varepsilon_{\mathrm{FG}} = 10^{-6}$. This choice represents an idealized level of foreground removal and thus the upper limit of the parameter constraints that 21\,cm IM experiments could achieve under favorable conditions. \cite{Wu:2021vfz} investigated the impact of different values of $\varepsilon_{\mathrm{FG}}$ on cosmological parameter constraints, showing that $\varepsilon_{\mathrm{FG}} \lesssim 10^{-5}$ is required to reliably extract the \(\mathrm{H\,{\scriptscriptstyle I}}\) signal and that increasing $\varepsilon_{\mathrm{FG}}$ from $10^{-6}$ to $10^{-5}$ degrades the associated parameter uncertainties by several tens of percent while leaving the overall qualitative conclusions robust.
} The parameters $\{A_X, n_X, m_X\}$ at $l_p=1000$ and $\nu_p=130$\,MHz are listed in Table~\ref{table:foreground_parameters}.

\begin{table}[t]
\caption{Foreground model parameters at $l_p = 1000$ and $\nu_p = 130$ MHz \citep{Santos_2005}.}
\label{table:foreground_parameters}
\centering
\begin{tabular}{lccc}
\hline\hline
Foreground & $A_X$ [mK\(^2\)] & $n_X$ & $m_X$ \\
\hline
Extragalactic point sources & 57.0  & 1.1 & 2.07 \\
Extragalactic free--free    & 0.014 & 1.0 & 2.10 \\
Galactic synchrotron        & 700   & 2.4 & 2.80 \\
Galactic free--free         & 0.088 & 3.0 & 2.15 \\
\hline
\end{tabular}
\end{table}

\begin{table*}[htbp]
  \captionsetup{justification=raggedright, singlelinecheck=false}
  \caption{Instrument parameters for the 21\,cm IM forecast.}
  \centering
  \renewcommand{\arraystretch}{1.15}
  \begin{adjustbox}{max width=\textwidth}
  \begin{tabular}{lcccccccc}
    \hline\hline
    Instrument & $z_{\min}$ & $z_{\max}$ & $N_\mathrm{d}$ & $N_\mathrm{b}$ & $D_\mathrm{dish}$ [m] & $S_{\text{area}}$ [deg$^2$] & $t_{\text{tot}}$ [h] & $T_{\text{inst}}$ [K] \\
    \hline
    QTT                             & 0 & 1        & 1  &96 & 110 & 20000 & 10000 & 20 \\
    JRT                             & 0 & 0.4 or 1        & 1  & 96 & 120 & 20000 & 10000 & 20 \\
    HRT                             & 0 & 0.4 or 1 & 1  & 96 & 120 & 20000 & 10000 & 20 \\
    \hline
  \end{tabular}
  \end{adjustbox}
  \label{table:instrument}
  \footnotesize
\end{table*}

\subsection{Instrument parameters}
\label{sec2.4}

In this paper, we forecast the performance of several large single-dish radio telescopes for 21\,cm IM. Table~\ref{table:instrument} summarizes the instrument parameters, redshift coverage \((z_{\min}, z_{\max})\), number of dishes \(N_\mathrm{d}\), beams per dish \(N_\mathrm{b}\), dish diameter \(D_\mathrm{dish}\), survey area \(S_\mathrm{area}\), total observing time \(t_\mathrm{tot}\), and receiver temperature \(T_\mathrm{inst}\). 

QTT is expected to be equipped with a PAF and a 0.7--1.8\,GHz receiver, corresponding to a redshift coverage of $0 \le z \le 1$ for the 21\,cm line. Following \cite{QTTP}, we set the number of beams to $N_\mathrm{b}=96$ and the system temperature to $T_\mathrm{inst}=20\,\mathrm{K}$. We assume $S_\mathrm{area}=2\times10^4\,\mathrm{deg}^2$ and $t_\mathrm{tot}=10^4\,\mathrm{h}$.
\begin{figure}[t]
       \centering
        \includegraphics[width=0.5\textwidth]{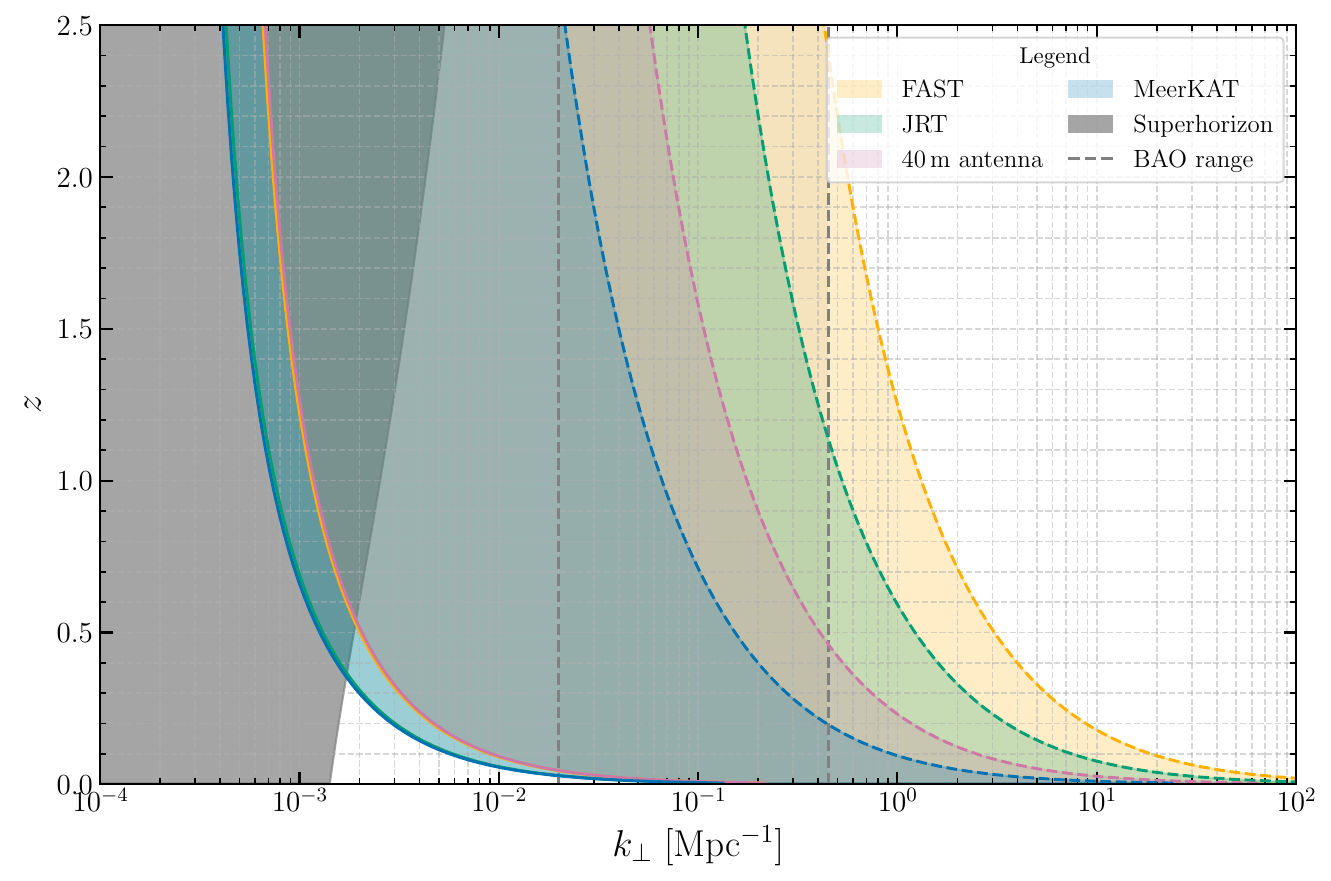}
          \caption{Redshift evolution of the transverse wavenumber $k_\perp$ coverage for different instruments, with colored shaded regions showing the minimum and maximum $k_\perp$ accessible at each redshift, dark gray dashed lines marking the BAO $k_\perp$ range and the shaded gray region denoting superhorizon scales with $k_H = 2\pi / r_H$.}

        \label{fig:krang}
\end{figure}
JRT is also expected to be equipped with a PAF and a 1--7\,GHz receiver~\citep[see][]{JRT:}, covering a redshift range of $0 \le z \le 0.4$ for 21\,cm. As the number of beams is not yet publicly available, we assume the same PAF configuration as for QTT, with $N_\mathrm{b}=96$ and $T_\mathrm{inst}=20\,\mathrm{K}$. The other parameters are also assumed to be $S_\mathrm{area}=2\times10^4\,\mathrm{deg}^2$ and $t_\mathrm{tot}=10^4\,\mathrm{h}$. 

Public information on HRT mainly concerns its construction scale, with no detailed specifications available for the PAF or other instrumental parameters. Therefore, we adopt the same settings as for JRT, including $N_\mathrm{b}=96$, $S_\mathrm{area}=2\times10^4\,\mathrm{deg}^2$, $t_\mathrm{tot}=10^4\,\mathrm{h}$, $T_\mathrm{inst}=20\,\mathrm{K}$, and a redshift range of $0 \le z \le 0.4$. {This assumption may introduce uncertainties in the predicted sensitivity and BAO/RSD measurement, as variations in the actual receiver temperature, beam count, or frequency coverage could shift the noise level and accessible modes. Once detailed specifications become publicly released, the forecasts can be updated by replacing the assumed HRT parameters with the official values.}

For the configurations with $z_{\max}$ values indicated as 0.4 or 1.0 in Table~\ref{table:instrument}, we perform forecasts for both upper limits. 
The value $z_{\max}=0.4$ corresponds to the currently planned design frequency coverage of JRT. 
To assess the potential scientific gains from extending the observing capability to higher redshift, we also consider $z_{\max}=1.0$. 
The latter corresponds to lowering the operating frequency to approximately 700--1000\,MHz, covering the range  $0.4 \le z \le 1$, which can substantially enlarge the accessible comoving volume, reduce cosmic variance, and improve the constraints on BAO parameters. 

Figure~\ref{fig:krang} shows the redshift evolution of the minimum and maximum transverse wavenumbers accessible to different instruments, represented by the shaded regions. The solid line corresponds to the survey-area limit where $k_{\perp}^{\min}= 2\pi / \sqrt{r^2 S_{\text{area}}}$, and the dashed line corresponds to the beam-resolution limit where $k_{\perp}^{\max}=2\pi D_{\text{dish}} / (r\lambda)$. A 40\,m-class telescope can cover the BAO wiggles up to redshift $z\simeq0.5$. At such low redshifts, the corresponding survey volume is small and the cosmic variance remains large, which limits its ability to constrain cosmological parameters, especially in $w_0w_a$CDM model. The 300\,m FAST telescope can resolve the transverse BAO modes easily over the redshift range $0<z<2.5$, but its fixed spherical reflector restricts sky accessibility, and its very narrow beam leads to a small instantaneous field of view. As a result, its effective mapping efficiency per unit observing time is low. 
In contrast, a fully steerable hundred-meter-class dish such as QTT or JRT provides an excellent balance between angular resolution and survey efficiency. It has sufficient resolution to cover all BAO wiggles up to redshift $z\simeq1$, while its beam is not too small to limit mapping speed. With its steerable design and the potential use of multi-beam or PAF, it can efficiently conduct wide-area surveys. Such hundred-meter-class telescopes are therefore highly suitable for 21\,cm cosmology.
Therefore, in the following analysis, we extend the 21\,cm observation capability of these instruments to $z_\mathrm{max}=1$.

In addition to analyzing each instrument individually, we also consider a combined survey configuration involving all three instruments, referred to as QTT+JRT+HRT. We assume that all three instruments observe the same sky area of \(20000\,\mathrm{deg}^2\), with identical redshift binning and frequency channel width. Each instrument observes independently for \(10^4\,\mathrm{h}\), and their thermal noise is assumed to be uncorrelated. The total noise is combined at the covariance level using inverse-variance weighting, as described in Equations~(\ref{eq:ct1}) and~(\ref{eq:ct2}). All instruments adopt the same foreground residual model \(C^\mathrm{F}\), which is treated as a common component and therefore not suppressed by cross-instrument averaging. If an instrument has a broader redshift coverage than the analysis range, we only include its data within the relevant redshift bins.

\subsection{Fisher matrix}
\label{sec2.5}

We construct the total covariance for each redshift bin and for each Fourier mode \(\boldsymbol{q},y\) by combining the cosmological signal \(C^{\mathrm{S}}\), thermal noise \(C^{\mathrm{N}}\), and residual foregrounds \(C^{\mathrm{F}}\).
For a single-dish, or for an array of identical dishes that observe the same sky area with the same instrumental response such as HRT and JRT, we take
\begin{equation} \label{eq:ct1}
C^{\mathrm{tot}}[\boldsymbol{q},y]
= C^{\mathrm{S}}[\boldsymbol{q},y]
+ C^{\mathrm{N}}[\boldsymbol{q},y]
+ C^{\mathrm{F}}[\boldsymbol{q},y].
\end{equation}
When combining two heterogeneous instruments that observe the same sky region, such as QTT and JRT, we assume uncorrelated thermal noise between instruments. The combined noise is added in inverse variance at the covariance level, yielding
\begin{equation}\label{eq:ct2}
C^{\mathrm{tot}}[\boldsymbol{q},y]
= C^{\mathrm{S}}[\boldsymbol{q},y]
+ \Bigl\{\,[C^{\mathrm{N}}_{1}]^{-1}[\boldsymbol{q},y] + [C^{\mathrm{N}}_{2}]^{-1}[\boldsymbol{q},y]\,\Bigr\}^{-1}
+ C^{\mathrm{F}}[\boldsymbol{q},y],
\end{equation}
which can be straightforwardly extended to more than two instruments.
Given the total covariance \(C^{\mathrm{tot}}\), the Fisher matrix for a set of parameters \(\{p_i\}\) is
\begin{equation}
F_{ij}
= \frac{1}{2}\,U_{\mathrm{bin}}
\int \frac{\mathrm{d}^{2}\!q\,\mathrm{d}y}{8\pi^{3}}
\left[
\frac{\partial \ln C^{\mathrm{tot}}[\boldsymbol{q},y]}{\partial p_{i}}\,
\frac{\partial \ln C^{\mathrm{tot}}[\boldsymbol{q},y]}{\partial p_{j}}
\right],
\end{equation}
where \(U_{\mathrm{bin}}=S_{\mathrm{area}}\,\Delta\tilde{\nu}\) is the survey volume factor for the redshift bin, and the integral extends over the accessible modes after applying the instrumental responses. In this work we treat for each redshift bin the BAO and RSD parameters \(\{D_{\mathrm{A}}(z),\,H(z),\,f\sigma_{8}(z)\}\) as the primary parameters. Here $D_{\mathrm{A}}(z)$ and $H(z)$ encode transverse and radial geometry, \(f\) is the linear growth rate, and \(\sigma_{8}\) is the rms matter fluctuation within spheres of radius \(8\,h^{-1}\,\mathrm{Mpc}\), with $h$ the dimensionless Hubble constant.

\subsection{Cosmological parameter constraints}
\label{sec2.6}

We use a Gaussian likelihood method to propagate the Fisher errors on the BAO/RSD parameters ${D_{\mathrm{A}}(z), H(z), f\sigma_{8}(z)}$ to cosmological parameters, with the covariance matrix obtained from the Fisher calculation. The joint posterior is then explored with MCMC using \texttt{Cobaya} package to obtain parameter constraints~\citep{Witzemann_2018, cobaya}. Our main parameter set is \(\{H_{0},\,\Omega_{\mathrm{m}},\,\Omega_{\mathrm{b}}, w_{0},\,w_{a}\}\). We assume spatial flatness with \(\Omega_\mathrm{K}=0\) and fix the total neutrino mass to \(\Sigma m_{\nu}=0.06\,\mathrm{eV}\). Uniform priors are adopted as follows: \(H_{0}\in[50,80]\,\mathrm{km\,s^{-1}\,Mpc^{-1}}\), \(\Omega_{\mathrm{m}}\in[0,0.7]\), \(\Omega_{\mathrm{b}}\in[0.03,0.07]\),  \(w_{0}\in[-3,5]\), \(w_{a}\in[-5,5]\).
Dark energy is described by an equation-of-state parameter \(w \equiv p/\rho\). We consider the \(\Lambda\)CDM and \(w_{0}w_{a}\)CDM models.
Furthermore, for comparison with current measurements, we also derive constraints from the DESI DR2 BAO data~\citep{DESI1}.

\section{Results and discussion}
\label{sec3}

\subsection{Present performance with current designs}

\begin{figure}[h]
\centering
\includegraphics[width=\hsize]{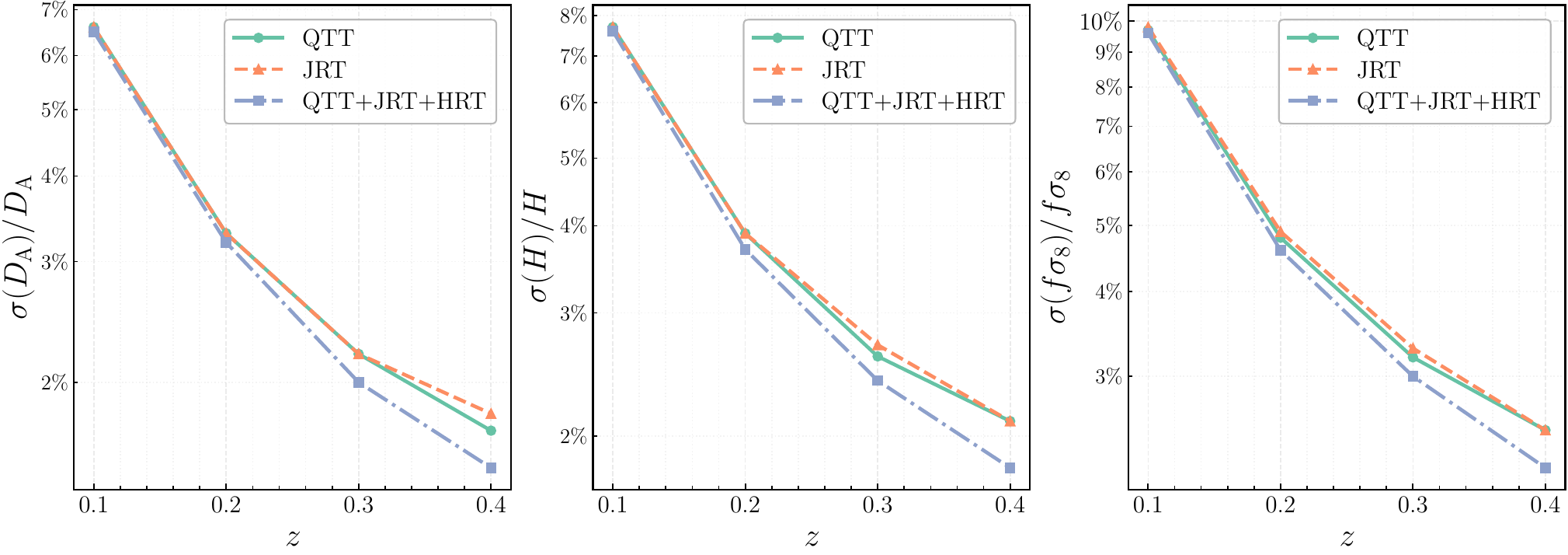}
\caption{Relative errors on $D_\mathrm{A}(z)$, $H(z)$, and $f\sigma_8(z)$ for 21-cm IM with QTT, JRT, and QTT+JRT+HRT for $z\in[0,0.4]$ using the current designs.}
\label{fig:0.4}
\end{figure}

\begin{figure}[h]
\centering
\includegraphics[width=0.48\hsize]{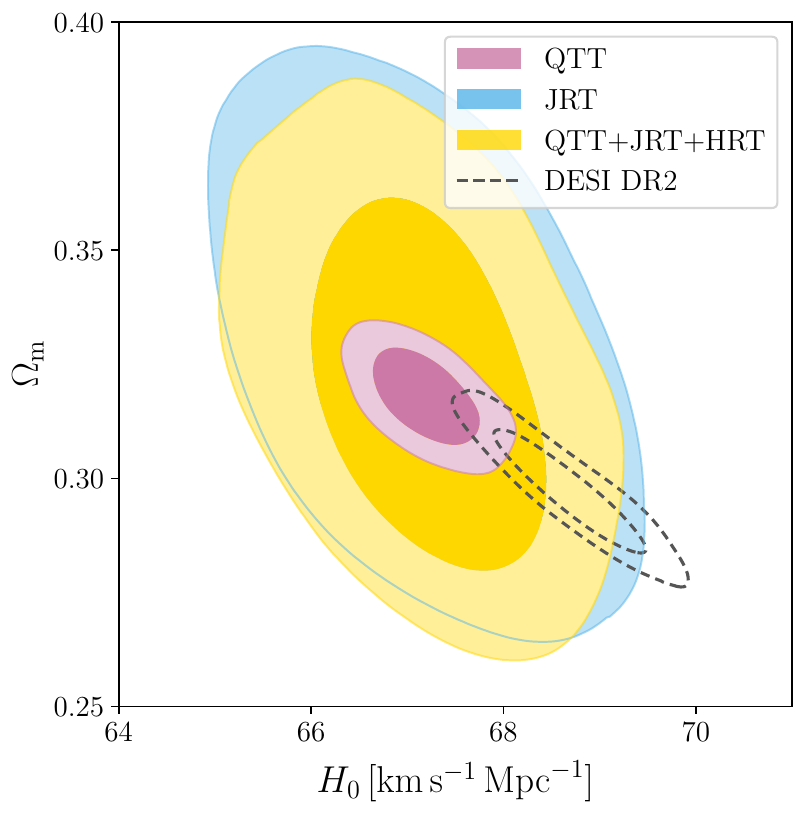}
\includegraphics[width=0.48\hsize]{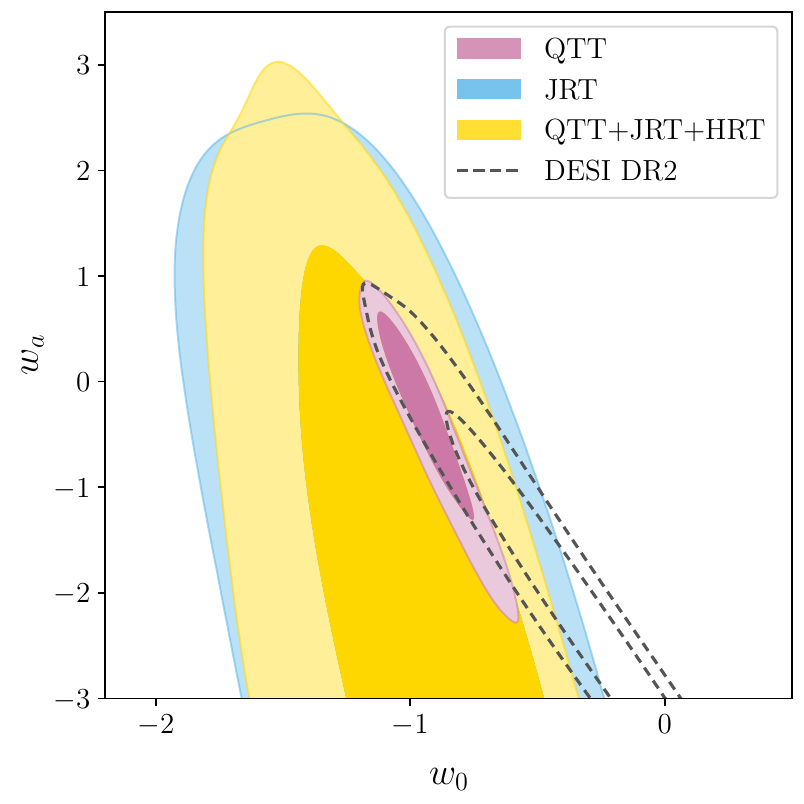}
\caption{$1\sigma$ and $2\sigma$ constraints on $\Lambda$CDM (left) and $w_0w_a$CDM (right) from simulated 21-cm IM with QTT, JRT, and QTT+JRT+HRT (restricted to $z\le0.4$), compared with DESI DR2.}
\label{fig:con0.4}
\end{figure}

\begin{table*}[t]
\centering
\caption{The $1\sigma$ uncertainties of the cosmological parameters in the $\Lambda$CDM and $w_0w_a$CDM models from DESI DR2 and 21\,cm IM, where $H_0$ is in units of km\,s$^{-1}$\,Mpc$^{-1}$.}
\label{tab:combined_sigma_only_compact_noexp}
\setlength{\tabcolsep}{3pt}
\renewcommand{\arraystretch}{1.12}

\footnotesize
\begin{tabular*}{0.8\linewidth}{@{\extracolsep{\fill}}
>{\raggedright\arraybackslash}p{3.2cm}
S[table-format=1.3, table-number-alignment=left] 
S[table-format=1.5, table-number-alignment=left] 
S[table-format=1.3, table-number-alignment=left] 
S[table-format=1.4, table-number-alignment=left] 
S[table-format=1.3, table-number-alignment=left] 
S[table-format=1.3, table-number-alignment=left] 
@{}}
\toprule
\multirow{2}{*}{$z$ range / Instrument} &
\multicolumn{2}{c}{$\Lambda$CDM} &
\multicolumn{4}{c}{$w_0w_a$CDM} \\
\cmidrule(lr){2-3}\cmidrule(lr){4-7}
& \multicolumn{1}{c}{$\sigma(H_0)$} &
\multicolumn{1}{c}{$\sigma(\Omega_\mathrm{m})$} &
\multicolumn{1}{c}{$\sigma(H_0)$} &
\multicolumn{1}{c}{$\sigma(\Omega_\mathrm{m})$} &
\multicolumn{1}{c}{$\sigma(w_0)$} &
\multicolumn{1}{c}{$\sigma(w_a)$} \\
\midrule
\multicolumn{7}{l}{$\bm{z\in[0, 0.4]}$} \\
\quad JRT & 0.936 & 0.02720 & 2.340 & 0.1030 & 0.724 & 6.400 \\
\quad QTT+JRT+HRT & 0.803 & 0.02610 & 2.210 & 0.0950 & 0.698 & 6.140 \\
\midrule
\addlinespace[0.3em]
\multicolumn{7}{l}{$\bm{z\in[0, 1]}$} \\
\quad QTT & 0.366 & 0.00688 & 1.010 & 0.0233 & 0.123 & 0.637 \\
\quad JRT & 0.346 & 0.00629 & 0.998 & 0.0237 & 0.116 & 0.610 \\
\quad QTT+JRT+HRT & 0.276 & 0.00477 & 0.814 & 0.0192 & 0.094 & 0.487 \\
\midrule
\addlinespace[0.3em]
\multicolumn{7}{l}{$\bm{z\in[0.3, 2.3]}$} \\
\quad DESI DR2 & 0.524 & 0.00897 & 3.250 & 0.0487 & 0.448 & 1.570 \\
\bottomrule
\end{tabular*}
\end{table*}

Figure~\ref{fig:0.4} presents the fractional errors on $D_{\mathrm A}(z)$, $H(z)$, and $f\sigma_8(z)$ for QTT, JRT, and QTT+JRT+HRT. QTT and JRT are nearly identical across redshift bins because they differ only slightly in aperture size. QTT+JRT+HRT provides the smallest errors at every redshift. 

Within $0.1 \le z \le 0.4$, errors decrease with redshift as the comoving volume per bin grows rapidly, increasing the number of independent Fourier modes. Near $z \simeq 0.1$, $D_{\mathrm A}(z)$, $H(z)$, and $f\sigma_8(z)$ are around $7\%$ to $10\%$ for all configurations. By $z \simeq 0.4$, QTT and JRT reach about $2\%$, and QTT+JRT+HRT yields smaller errors.
These trends follow directly from Equations~\eqref{eq:cs} and~\eqref{eq:cn}. With \(t_{\mathrm{tot}}\) and \(\Delta\nu\) held fixed, increasing \(N_{\mathrm{b}}N_{\mathrm{d}}\) reduces the per-mode thermal noise. Increasing the survey area \(S_{\mathrm{area}}\) raises the per-mode noise through \(U_{\mathrm{bin}}\) because the dwell time per solid angle decreases. At the same time, a larger area provides more independent modes, which reduces the overall statistical uncertainty in the Fisher analysis.

Figure~\ref{fig:con0.4} shows the constraints on the $\Lambda$CDM and $w_0w_a$CDM model parameters. These constraints are derived from the current design configurations of QTT, JRT, and QTT+JRT+HRT at the $1\sigma$ and $2\sigma$ confidence levels. For clarity, in the combined case QTT+JRT+HRT we restrict the analysis to the overlapping design range $0\le z\le 0.4$ to reflect the present JRT and HRT frequency coverage, and the combined case does not include QTT data from $z=0.4$ to $1$. In the $\Lambda$CDM model, for QTT, JRT, and QTT+JRT+HRT, $\sigma(H_0)$ is $0.366$, $0.936$, and $0.803~\mathrm{km\,s^{-1}\,Mpc^{-1}}$, respectively, and $\sigma(\Omega_{\mathrm m})$ is 0.00688, 0.02720, and 0.02610, respectively. In the $w_0w_a$CDM model, the uncertainties on $w_0$ are $\sigma(w_0)=0.123$ for QTT, $0.724$ for JRT, and $0.698$ for QTT+JRT+HRT. For $w_a$, the corresponding uncertainties are $\sigma(w_a)=0.637$, $6.400$, and $6.140$, respectively.

Under these current designs, QTT already achieves tighter constraints than DESI DR2. In the $\Lambda$CDM model QTT gives $\sigma(H_0)=0.366~\mathrm{km\,s^{-1}\,Mpc^{-1}}$, smaller than 0.524 from DESI DR2. In the $w_0w_a$CDM model QTT yields $\sigma(w_0)=0.123$ and $\sigma(w_a)=0.637$, both smaller than the DESI DR2 results 0.448 and 1.570. In contrast, the limited redshift range $0 \le z \le 0.4$ of JRT, HRT, and QTT+JRT+HRT leads to weaker performance in both models.

\subsection{Impact of extending the redshift coverage to $z_{\max}=1$}

\begin{figure}[t]
\centering
\includegraphics[width=\hsize]{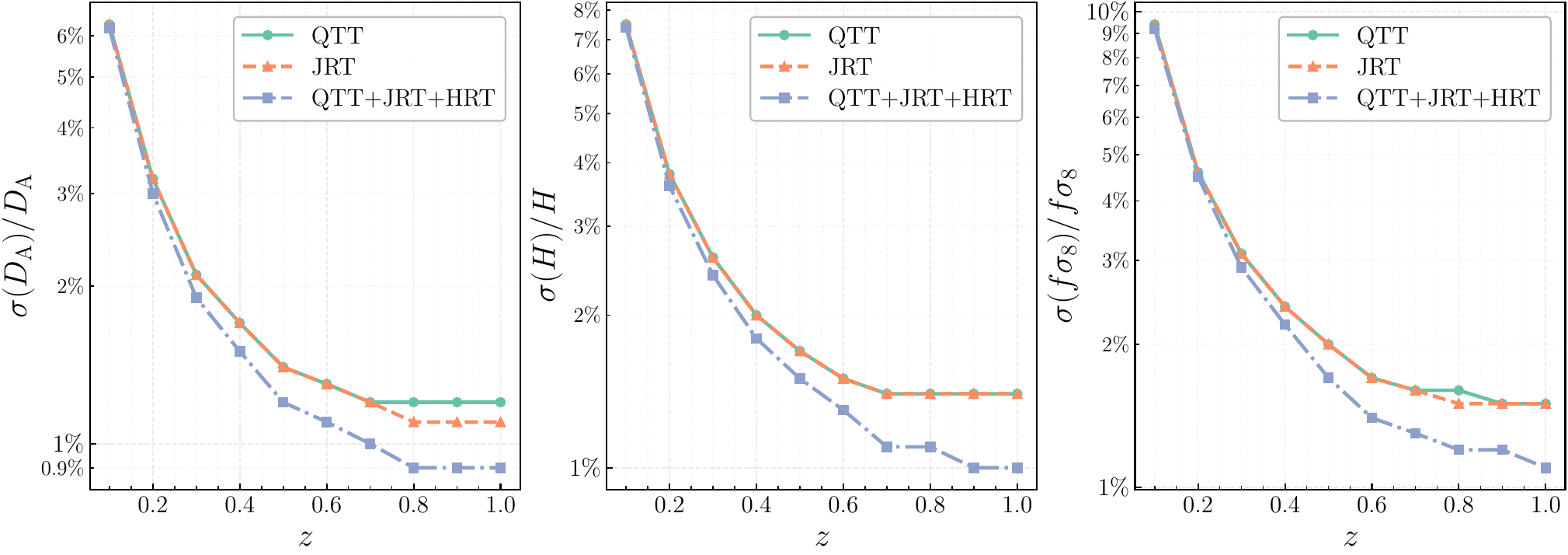}
\caption{Relative errors on $D_\mathrm{A}(z)$, $H(z)$, and $f\sigma_8(z)$ for 21-cm IM with QTT, JRT, and QTT+JRT+HRT for $z\in[0,1]$, assuming JRT and HRT are extended to $z_{\max}=1$.}
\label{fig:1}
\end{figure}

\begin{figure}[t]
\centering
\includegraphics[width=0.48\hsize]{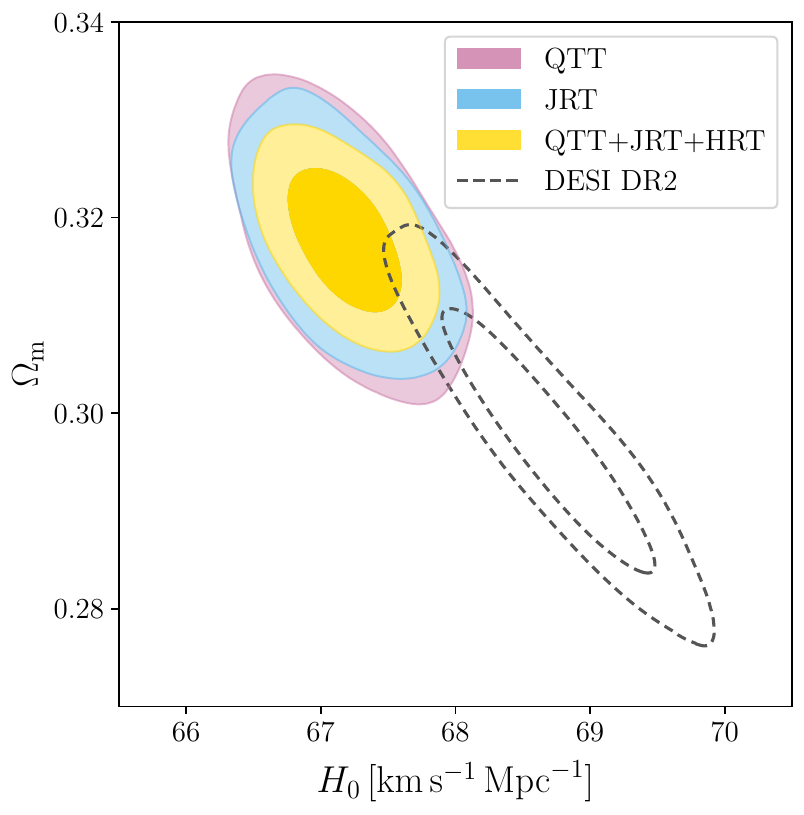}
\includegraphics[width=0.48\hsize]{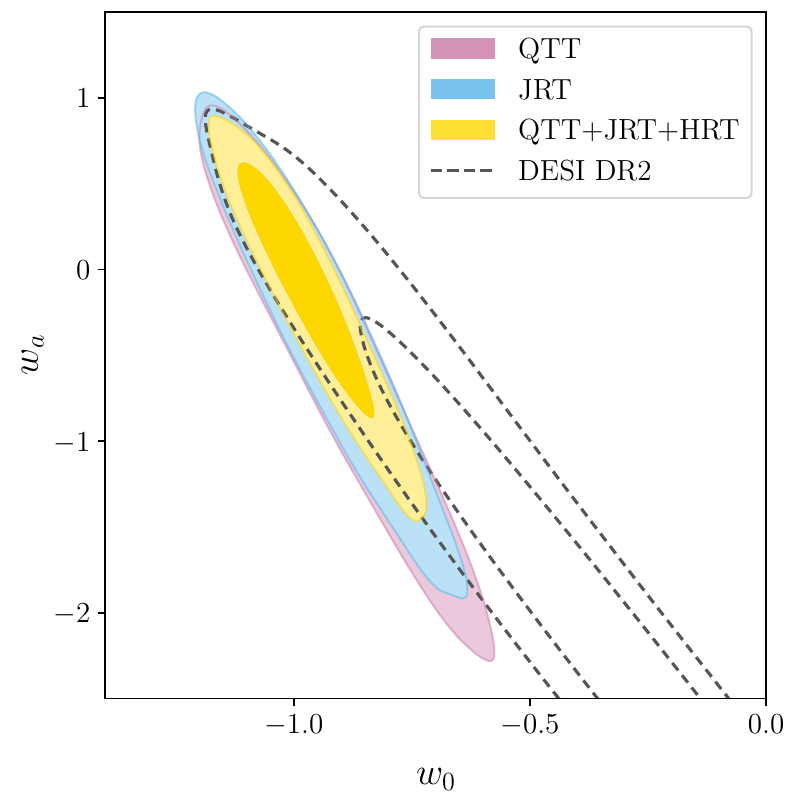}
\caption{$1\sigma$ and $2\sigma$ constraints on $\Lambda$CDM (left) and $w_0w_a$CDM (right) from simulated 21-cm IM with QTT, JRT, and QTT+JRT+HRT for the extended configuration with $z_{\max}=1$, compared with DESI DR2.}
\label{fig:con1}
\end{figure}

Building on the previous discussion, we further investigate the improvement from extending the redshift coverage of both JRT and HRT up to $z_{\mathrm{max}} = 1$.
Figure~\ref{fig:1} presents the fractional errors in $D_{\mathrm A}(z)$, $H(z)$, and $f\sigma_8(z)$ from $z=0$ to $z=1$ for QTT, JRT, and QTT+JRT+HRT. The instrument performance ranking established at low redshift remains consistent across the extended redshift range. QTT+JRT+HRT is most precise at all redshifts. QTT and JRT exhibit nearly identical performance, with their curves overlapping across most redshift bins.

In the $\Lambda$CDM model, JRT reduces $\sigma(H_0)$ from $0.936~\mathrm{km\,s^{-1}\,Mpc^{-1}}$ to $0.346~\mathrm{km\,s^{-1}\,Mpc^{-1}}$, corresponding to a decrease of approximately 63.0\%. For QTT+JRT+HRT, $\sigma(H_0)$ is lowered from $0.803~\mathrm{km\,s^{-1}\,Mpc^{-1}}$ to $0.276~\mathrm{km\,s^{-1}\,Mpc^{-1}}$, marking a reduction of about 65.6\%. In the $w_0w_a$CDM model, $\sigma(w_0)$ for JRT decreases from 0.724 to 0.116, corresponding to a reduction of 83.9\%. Meanwhile, $\sigma(w_a)$ declines from 6.40 to 0.610, representing a reduction of 90.5\%. For QTT+JRT+HRT, $\sigma(w_0)$ decreases from 0.698 to 0.094, and $\sigma(w_a)$ from 6.14 to 0.487, yielding reductions of 86.5\% and 92.1\%, respectively. Compared with the current optical results from DESI DR2, all instruments provide significantly tighter cosmological parameter constraints. Relative to DESI DR2, QTT+JRT+HRT lowers $\sigma(H_0)$ by 47.3\%, while QTT and JRT show decreases of 30.2\% and 34.0\% in the $\Lambda$CDM model. In the $w_0w_a$CDM model, QTT+JRT+HRT yields $\sigma(w_0)$ and $\sigma(w_a)$ that are lower by 79.0\% and 69.0\%, respectively.
For QTT and JRT, the reductions are 72.5\% and 74.1\% in $\sigma(w_0)$, and 59.4\% and 61.1\% in $\sigma(w_a)$.

Overall, extending the upper redshift limit to \(z_{\mathrm{max}} = 1\) increases the comoving volume and the number of independent modes, allowing all instruments to achieve tighter constraints than DESI DR2. This demonstrates the strong potential of coordinated 21\,cm IM instruments for measuring dark energy EoS parameters.

\section{Conclusions}
\label{sec4}

In this work, we evaluate the cosmological performance of 21\,cm IM for QTT, JRT, and HRT. We adopt a semi-analytic Fisher-matrix framework to forecast constraints on the BAO/RSD parameters $D_\mathrm{A}(z)$, $H(z)$, and $f\sigma_8(z)$ across a range of redshifts, and propagate these constraints into parameter forecasts for the $\Lambda$CDM and $w_0w_a$CDM models.

Under these current designs, QTT already delivers tighter constraints than DESI DR2 for both $\Lambda$CDM and $w_0w_a$CDM, benefiting from its wider redshift coverage and multi-beam capability. In contrast, JRT and HRT, as well as the
 QTT+JRT+HRT restricted to $0\le z\le0.4$, are limited by the smaller effective volume and fewer redshift bins, and therefore yield weaker constraints, especially on $w_0w_a$CDM model. Extending the coverage to $z_{\max}=1$ markedly improves performance. The larger comoving volume and the increased number of independent Fourier modes reduce statistical uncertainties of the BAO/RSD parameters. In this configuration, \textsc{QTT+JRT+HRT} achieves $\sim\!1\%$ relative errors on the BAO parameters at intermediate redshift and attains the constraints of $\sigma(w_0)=0.094$ and $\sigma(w_a)=0.487$, outperforming current optical results from DESI DR2.

In summary, these hundred-meter–class single-dish telescopes are well suited for 21\,cm IM cosmology, particularly for constraining dark energy EoS parameters. Extending their redshift coverage would further enhance their scientific potential. Our work offers practical guidance for future survey design to fully realize the cosmological capabilities of 21\,cm IM.

\begin{acknowledgements}
We thank Peng-Ju Wu for helpful discussions. This work was supported by the National SKA Program of China (Grants Nos. 2022SKA0110200 and 2022SKA0110203), the National Natural Science Foundation of China (Grants Nos. 12533001, 12575049, 12473001, and 12473091), the China Manned Space Program (Grant No. CMS-CSST-2025-A02), the National 111 Project (Grant No. B16009), and the Fundamental Research Funds for the Central Universities (Grant No. N2405008).
\end{acknowledgements}

\bibliographystyle{aa}
\bibliography{aa}

\end{document}